\documentclass[10pt,letterpaper]{article}
\usepackage[top=0.85in,left=2.75in,footskip=0.75in,marginparwidth=2in]{geometry}

\usepackage[utf8]{inputenc}

\usepackage{cite}

\usepackage{nameref,hyperref}

\usepackage[right]{lineno}

\usepackage{microtype}
\DisableLigatures[f]{encoding = *, family = * }

\raggedright
\usepackage{changepage}

\usepackage[aboveskip=1pt,labelfont=bf,labelsep=period,singlelinecheck=off]{caption}

\makeatletter
\renewcommand{\@biblabel}[1]{\quad#1.}
\makeatother

\usepackage{lastpage,fancyhdr,graphicx}
\usepackage{epstopdf}
\fancyheadoffset[L]{2.25in}
\fancyfootoffset[L]{2.25in}

\usepackage{color}

\definecolor{Gray}{gray}{.25}

\usepackage{graphicx}

\usepackage{sidecap}

\usepackage{wrapfig}
\usepackage[pscoord]{eso-pic}
\usepackage[fulladjust]{marginnote}
\reversemarginpar

\usepackage{hyperref}
\usepackage{setspace}
\usepackage{dcolumn}
\usepackage{amsthm}
\usepackage{amssymb}
\usepackage{listings}
\usepackage{calc}
\usepackage{verbatim}
\usepackage{appendix}
\usepackage{mathtools}
\usepackage{booktabs}
\usepackage{blkarray}

\begin{document}
\vspace*{0.35in}

% title goes here:
\begin{flushleft}
{\Large
\textbf\newline{Cheater-altruist synergy in immunopathogenic ecological public goods games}
}
\newline
% authors go here:
\\
Bryce Morsky\textsuperscript{*},
Dervis Can Vural
\\
\bigskip
Department of Physics, University of Notre Dame, Nieuwland Science Hall, Notre Dame, 46556, USA
\\
\bigskip
* bryce.morsky@gmail.com

\end{flushleft}

\section*{Abstract}
Much research has focused on the deleterious effects of free-riding in public goods games, and a variety of mechanisms that suppresses cheating behaviour. Here we argue that under certain conditions cheating behaviour can be beneficial to the population. In a public goods game, cheaters do not pay for the cost of the public goods, yet they receive the benefit. Although this free-riding harms the entire population in the long run, the success of cheaters may aid the population when there is a common enemy that antagonizes both cooperators and cheaters. Here we study models in which an immune system antagonizes a cooperating pathogen. We investigate three population dynamics models, and determine under what conditions the presence of cheaters help defeat the immune system. The mechanism of action is that a polymorphism of cheaters and altruists optimizes the average growth rate. Our results give support for a possible synergy between cooperators and cheaters in ecological public goods games.

% now start line numbers

\section*{Introduction}
The sociology of microorganisms is an important and growing field of study \cite{west06}, and altruism and conflict, as applied to biofilm evolution, are important factors within its domain \cite{boyle15, xavier07}. Public goods benefit every agent, but more so the free-riders that use them without contributing. An example is siderophore production in bacterial population \cite{cordero12}. Iron is an import and scare resource for bacteria living in hosts. Thus, they produce siderophores that bind to iron in hemoglobin and other molecules to form iron-siderophore complexes. The bacteria then absorb these complexes. Some bacteria cheat, by not producing (or producing fewer) siderophores. They absorb the iron-siderophore complexes produced by the community as a whole, without contributing to the cost \cite{boyle13}. The lower operating cost allows cheating strains to reproduce faster, dominate the population, and lead to an iron-deprived community. There are several questions that arise from this social dilemma: how can altruists survive with competing cheaters? What role do cheaters play within the infection? If altruists and cheaters can coexist, how do they interact and evolve?

Ecological public goods games have been shown to facilitate cooperation where population density depends on average payoff \cite{hauert06,hauert08}. Other mechanisms proposed to facilitate altruism include kin selection \cite{hamilton63, queller92}, and reciprocal altruism \cite{ale13}. Experimental studies of altruistic siderophore production in \textit{Pseudomas aeruginosa} have shown that higher levels of cooperation are observed in higher relatedness communities. More localized competition selects for lower levels of siderophore production. Furthermore there is a significant interaction between relatedness and the scale of competition, with relatedness having less effect when the scale of competition is more local \cite{griffin04}. Studies of \textit{P.\ aeruginosa} have shown another solution, metabolic constraints on social cheating. Quorum sensing can control both these public goods (extracellular proteases) and private goods (cellular enzymes) \cite{dandekar12}. Additionally, it is well known that space can facilitate cooperation and coexistence of cooperators and defectors. This effect has been shown in both theoretical models \cite{roca09} and in experimental bacterial populations \cite{hol13,rainey03}. A review of altruism in microbial communities that explore a collection of these mechanism and issue is found in \cite{damore12}.

A variety of interesting pathogen-immune system models has been explored in the literature \cite{forys02, kirschner98, kuznetsov94, perelson97}. Here, we extend the model in \cite{kuznetsov94} by the incorporation of a public goods game, and interpret their model as a general host-pathogen model. We explore both linear and Monod public good growth function, and adapt two other canonical two-species growth models into our models that are adaptations of the logistic equation \cite{crow70}. Depending on the specific growth source used by the microbes, empirical data either supports a linear or Monod growth rate function \cite{monod49}. We find that the models employing the Monod function exhibit a synergy between altruists and cheaters where the public good is more efficiently used to increase the growth rate of the entire population of pathogens. With this effect, the pathogens can overcome the immune response of the host whereas wholly altruistic or cheating populations cannot.

\section*{Methods}
Let $x_a$ and $x_c$ be the numbers of altruists and cheaters, respectively, $X = x_a + x_c$ be the total pathogen population, and $y$ be the number of immune system effectors. The growth rate of the cheater pathogens, $r_c$, is the sum of the basal growth rate, $\beta$, and the benefit from the public good, $g(x_a / X)$, which is a function of the proportion of the population that are altruists. We explored two public good growth functions, linear (\ref{lingrowth}) and Monod growth (\ref{Mongrowth}):

\begin{align}
g \left( \frac{x_a}{X} \right) &= \frac{\alpha x_a}{X}, \label{lingrowth} \\
g \left( \frac{x_a}{X} \right) &= \frac{\alpha x_a/X}{K_{\alpha} + x_a/X}, \label{Mongrowth} 
\end{align}

\noindent where $\alpha$ is the maximum growth rate provided by the public good, and $K_{\alpha}$ is the half velocity constant. The growth rate for altruists is $r_a=r_c-c$, where $c$ is the cost of producing the public good. Let $K$ be the carrying capacity and $\bar{r} = (r_a x_a + r_c x_c)/X$ be the average rate of growth. In the absence of an immune response and other complications, it is standard to describe the growth and competition of bacteria by logistic dynamics\cite{crow70}. Here we separately consider three models in this class: $r$/$K$ selection (\ref{rKsel}), weak selection (\ref{weak}), and interspecific competition (\ref{intersp}).

\begin{align}
\dot{x}_a &= r_a x_a F_a(x_a,x_c), \\
\dot{x}_c &= r_c x_c F_c(x_a,x_c), \\
F_i(x_a,x_c)&= \left(1 - \frac{X}{K} \right), \label{rKsel} \\
F_i(x_a,x_c) &= \left(1 - \frac{X}{r_i K} \right), \label{weak} \\
F_i(x_a,x_c) &= \left(1 - \frac{\bar{r}X}{r_i K} \right). \label{intersp}
\end{align}

In $r$/$K$ selection (\ref{rKsel}), the success of one phenotype over the other is determined by both the growth rates and carrying capacities. There is a trade-off between $r_i$ and $K$: when close to the carrying capacity, $K$ determines selection. And, when the population is small, $r_i$ determines selection. In weak selection (\ref{weak}), the population is limited by $r_iK$, and this trade-off does not exist. Finally, in interspecific competition (\ref{intersp}), the phenotypes compete against one another for resources. A phenotype with a larger growth rate will curtail the carrying capacity of its competitor.

In our model, the immune system does not differentiate between cooperating and free riding pathogens, and thus $\dot{y}$ depends only on $X$. The immune system's effectors, which kill foreign cells, are produced at a basal rate, $\sigma$, and die at a rate, $\delta$. The pathogen free host thus has an equilibrium $y^* = \sigma/\delta$. In the presence of a pathogen, the production of immune agents is determined by the nonlinear activation function, $\rho yX/(\eta + X)$, and immune agents are exhausted at a rate, $\mu X$. The immune system attacks the pathogen, and reduces their number at a rate, $x_i y$. The dynamics of this system are governed by the following system of equations:

\begin{align}
\dot{x}_a &= r_a x_a F(x_a,x_c)  - x_a y, \\
\dot{x}_c &= r_c x_c F(x_a,x_c)  - x_c y, \\
\dot{y} &= y \left( \frac{\rho X}{\eta + X} - \mu X - \delta \right) + \sigma.
\end{align}

This model is based on \cite{kuznetsov94}, where it was shown to agree with empirical observations. Here we have generalized it to include bacteria with two kinds of social behaviour. A summary of the parameters, variables, and their values (from \cite{kuznetsov94}) can be found in Table \ref{param}. We have chosen the values of $\alpha$, $K_{\alpha}$, and $c$ in Table \ref{param} to highlight the synergistic behaviour of cheaters and altruists, since the synergy is not present throughout parameter space.

\begin{table}
\begin{center}
\begin{tabular}{ccl}
\toprule
\begin{tabular}{@{}c@{}}Parameter/ \\ variable\end{tabular} & Value & Definition \\
\midrule
$\alpha$ & $1.3088$ & Benefit from the public good \\
$\beta$ & $1.636$ & Basal growth rate \\
$\delta$ & $0.3743$ & Death rate of immune agents \\
$\eta$ & $20.19$ & Activation parameter \\
$K$ & $500$ & Carrying capacity \\
$K_{\alpha}$ & $0.5$ & Monod parameter \\
$\mu$ & $0.00311$ & Inactivation rate \\
$\rho$ & $1.131$ & Activation parameter \\
$\sigma$ & $0.1181$ & Birth rate of immune agents \\
$c$ & $0.818$ & Cost to altruists \\
$r_a$ & --- & Altruist growth rate \\
$r_c$ & --- & Cheater growth rate \\
$\bar{r}$ & --- & Average growth rate \\
$x_a$ & --- & No.\ of altruists \\
$x_c$ & --- & No.\ of cheaters \\
$X$ & --- & Total pathogen population \\
$y$ & --- & No.\ of immune agents \\
\bottomrule
\end{tabular}
\end{center}
\vspace{-2mm}
\caption{Summary definitions of parameters and variables.} \label{param}
\end{table}

\section*{Results}
Here we discuss the qualitative dynamics of immune system plus social and anti-social bacteria, i.e. the equilibria, stability, and invariant surfaces. We follow these analyses with simulation results that depict the synergy between altruists and cheaters, and the effects of different parameters on this synergy.

At equilibrium, we have $\dot{x}_a = \dot{x}_c$, which implies that $(r_c - c) F_a = r_c F_c$. In all three models, this reduces to $r_c - c = r_c$. Thus, assuming that $c>0$, there are no polymorphic pathogen equilibria. \ref{rK_interequil} and \ref{weakequil} are the equilibria at these monomorphic populations for $r$/$K$ selection and inter-specific competition, and weak selection, respectively;

\begin{align}
x_i &= K \left( 1 - \frac{y}{r_i} \right), \label{rK_interequil} \\
x_i &= K (r_i - y) \label{weakequil},
\end{align}

\noindent with $r_a = \beta + g(1) - c$ and $r_c = \beta$ (note that since $\bar{r} = r_i$ in a monomorphic population of phenotype $i$, \ref{rK_interequil} applies for both $r$/$K$ selection and interspecific competition). Since we assume that the public good is beneficial to the pathogens in a monomorphic altruistic population, $g(1)>c \implies r_a > r_c$, it can be easily seen by \ref{rK_interequil} and \ref{weakequil} that all monomorphic altruist equilibria have higher pathogen counts than their monomorphic cheater counterparts. However, in a polymorphic population, $r_c = \beta + g(x_a/X) > \beta + g(x_a/X) - c = r_a$. Since the growth rate for cheaters is always greater than the growth rate of altruists in such a case, there are no stable monomorphic altruistic equilibria.

\begin{figure}[!hb]
\includegraphics[width=\linewidth]{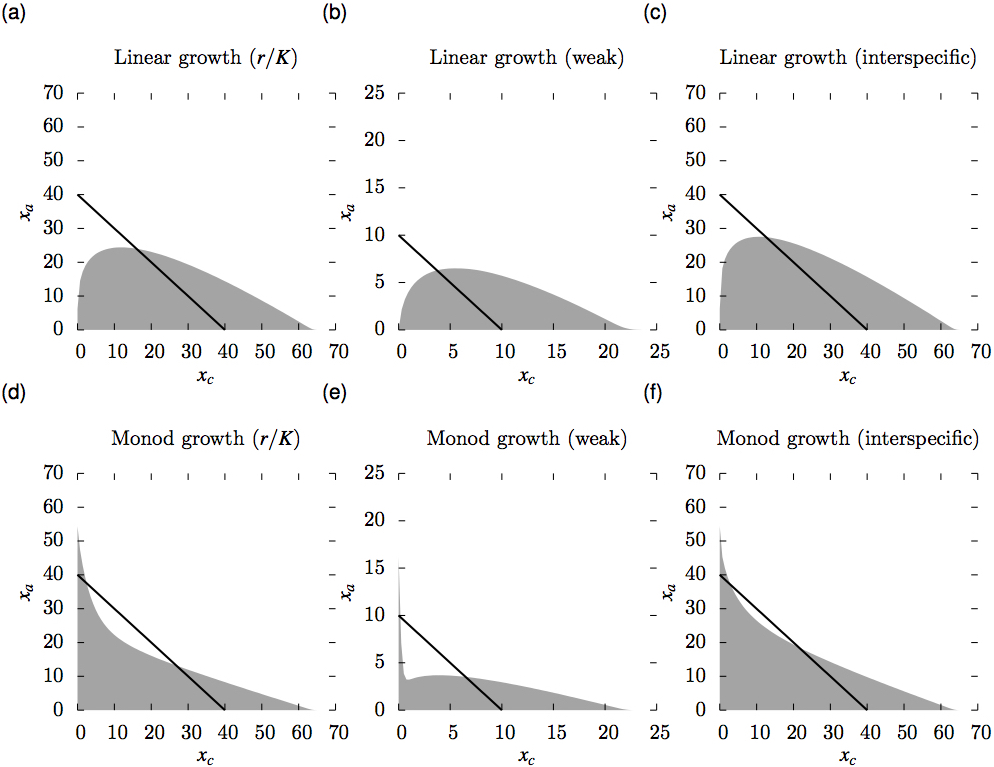}
\caption{A mixed population of altruists and cheaters minimizes the number of pathogens required to overcome the immune system when growth from the public good behaves as a Monod function (\textbf{d}-\textbf{f}). However, this behaviour is not observed when the growth function is linear (\textbf{a}-\textbf{c}). The white and gray regions are where the pathogens overcome and are suppressed by the immune system, respectively. The black curves are the isoclines, $x_a = X - x_c$, where $X=10$ in \textbf{b} and \textbf{e}, and $X=40$, otherwise.} \label{separatrices}
\end{figure}

The model has four fixed points: the pathogen free state, a suppressed population of pathogens (i.e. corresponds to a dormant state); a very large population of pathogens (i.e. bacteria take over the host); and a saddle point (i.e. the bacteria are populous, but have not taken over the host). The pathogen free state is connected to the saddle via a stable manifold. This stable manifold, a separatrix, divides phase space into regions where the pathogens succeed and fail. The unstable manifold spirals into the suppressed state on one side of the separatrix, and connects to the success state on the other side. Qualitatively, this picture is the same as in \cite{kuznetsov94} with the addition of the altruist dimension.

Fig.~\ref{separatrices} shows the regions of suppression (gray) and success (white) of pathogens for all the models we study. The goal here is to check whether a certain \emph{initial} population of microbes $(x_c, x_a)$ succeeds in defeating the host. The line $x_a = X - x_c$ (with constant $X$) is overlaid to these plots to show whether changing the composition of the population---without changing the total number of bacteria---results in a difference in the fate of the disease.

Interestingly, in some cases we observe that population compositions with an intermediate number of cheaters can place the population in the successful region, while too little or too many cheaters jeopardize the population. In other words, while neither pure cooperation nor pure cheating leads to success, a mixture of the two does. We observe this phenomenon in bacteria growing according to the Monod law, but not for linear growth.

We can explain this phenomenon by examining the equations for the change in the total population, $\dot{X}$, $r$/$K$ selection and interspecific competition (\ref{rK_intertotal}), and weak selection (\ref{weaktotal});

\begin{align}
\dot{X} &= \bar{r}X\left( 1 - \frac{X}{K} \right) - yX, \label{rK_intertotal} \\
\dot{X} &= \bar{r}X - \frac{X^{\mathrlap{2}}}{K} - yX. \label{weaktotal}
\end{align}

\noindent In linear growth, $\bar{r} = \beta + (\alpha - c)x_a/X$, which is an increasing function with respect to $x_a$ (given $\alpha > c)$. Thus, \ref{rK_intertotal} and \ref{weaktotal} are increasing with respect to $x_a$. Therefore, the impact of decreasing altruists in favor of cheaters is a decrease in the total population's rate of growth; cheaters harm the population as a whole. However, for Monod growth, we have the function

\begin{equation}
\bar{r} = \beta + \left( \frac{\alpha}{K_{\alpha} + x_a/X} - c \right)\frac{x_a}{X} \label{average_Monod}
\end{equation}

\noindent with the assumption that $\alpha/(K_\alpha + 1) > c$. With respect to the proportion of altruists, this function has a local maximum at

\begin{equation}
\frac{x_a}{X} = \sqrt{\frac{\alpha K_\alpha}{c}} - K_\alpha > \sqrt{K_\alpha (K_\alpha+1)} - K_\alpha > 0.
\end{equation}

\noindent Therefore, unlike the linear growth case, the optimal growth rate will occur in the presence of cheaters when $\sqrt{\alpha K_\alpha/c} - K_\alpha < 1$.

\begin{figure}[!ht]
\includegraphics[width=\linewidth]{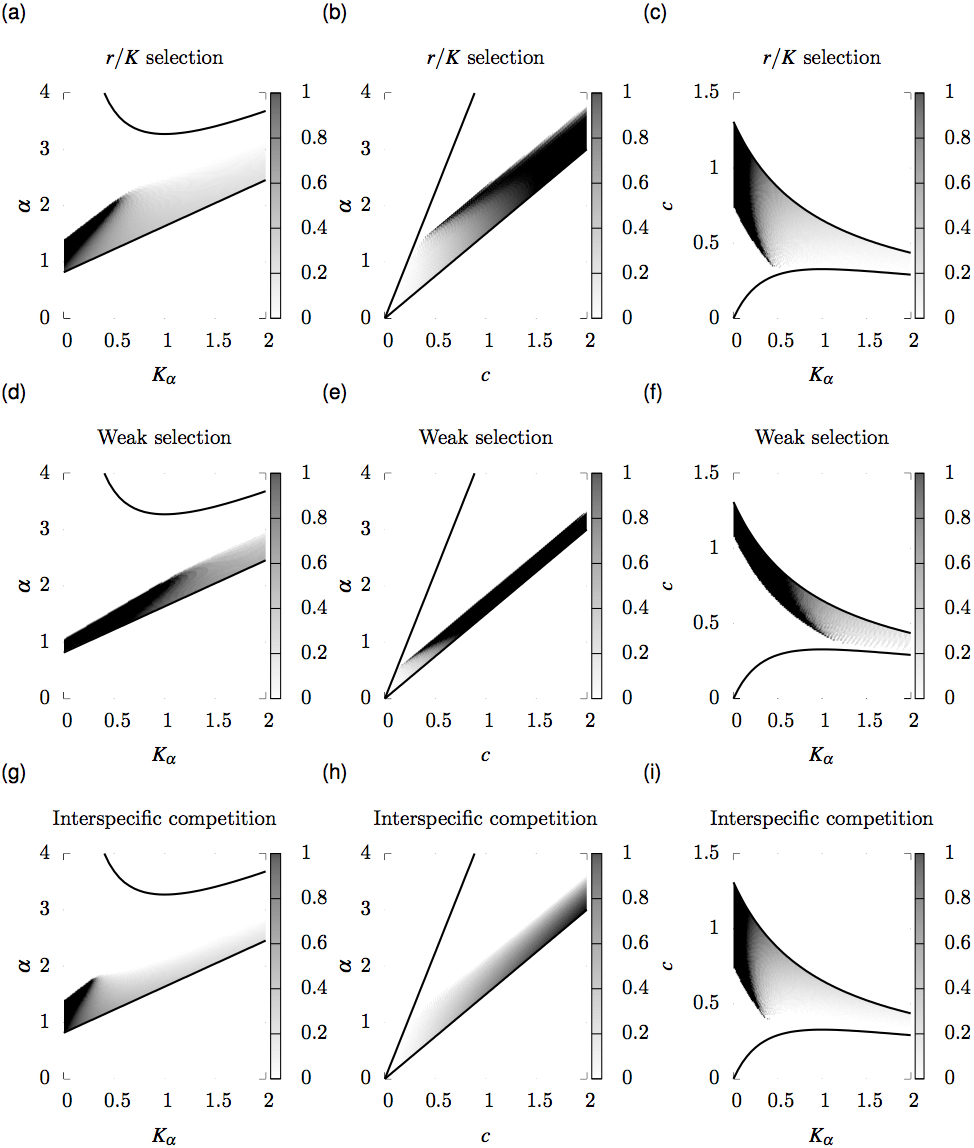}
\caption{Comparison of success of a $5\%$ cheater population vs.\ a pure altruist population for varying parameters $\alpha$, $K_{\alpha}$, and $c$. The gray scale measures the difference between the number of pathogens required to overcome immune suppression for the $5\%$ cheater, $X$, vs.\ pure altruist populations, $X^\prime$, divided by $X$. The curves define the envelope in which the optimal growth rate occurs for $x_a < 1$ and the pure altruist Monod growth rate is greater than $c$. Where a parameter is not varied, its value is from Table~\ref{param}.} \label{vary}
\end{figure}

Fig.~\ref{vary} shows the results for simulations where we varied the parameters, $\alpha$ (the maximum growth rate from the public good), $K_\alpha$ (the half velocity constant), and $c$ (the cost of public good production).  We ran simulations for populations of size $X$ with $5\%$ of the population cheaters. We incrementally increased $X$ until the pathogens succeeded. We then simulated a complete altruist population of size $X^\prime=X$. If this population was suppressed, we increased $X^\prime$ until it succeeded, and colored the graphs by magnitude $(X^\prime-X)/X$.

\begin{figure}[!ht]
\includegraphics[width=\linewidth]{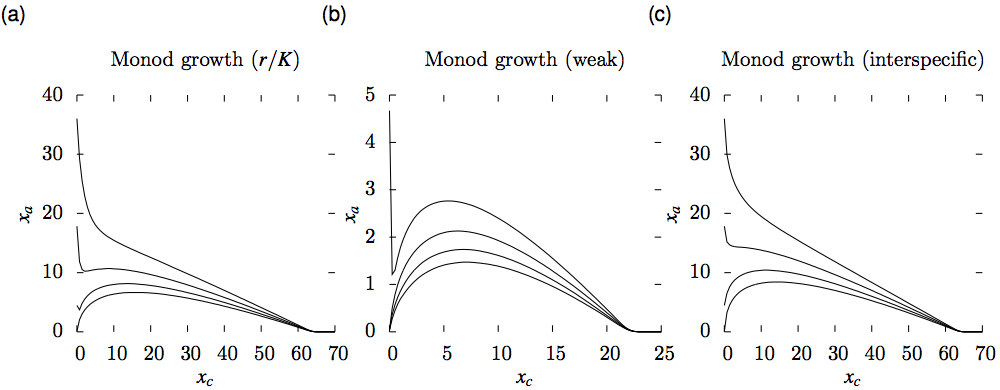}
\caption{Comparison of varying benefits from public good growth rates ($\alpha$). All curves are in decreasing order from top to bottom of the graphs with increasing $\alpha$ ($\alpha=1.5$, $1.75$, $2$, and $2.25$). As $\alpha$ increases, the behaviour of the model approaches that of linear growth, i.\ e.\ a monomorphic altruistic population is optimal with respect to the pathogens.} \label{alpha}
\end{figure}

We observe that for sufficiently large $\alpha$ and small $c$, cheaters do not benefit the population. In these cases, the public good is cheap and efficacious. However, within the region where we observe cheater-altruist synergy, increasing $\alpha$ intensifies the synergistic effect except with respect to $c$ in interspecific competition. Figure~\ref{alpha} explains this effect. Note the sharp drop in the separatrix in Figure~\ref{alpha}; a small proportion of altruists is beneficial to the pathogens. However, the remainder of the curve shows malign effects of increasing the proportion of cheaters. As we increase $\alpha$, this phenomenon disappears, and we observe the same qualitative behaviour as linear growth.

\begin{figure}[!hb]
\includegraphics[width=\linewidth]{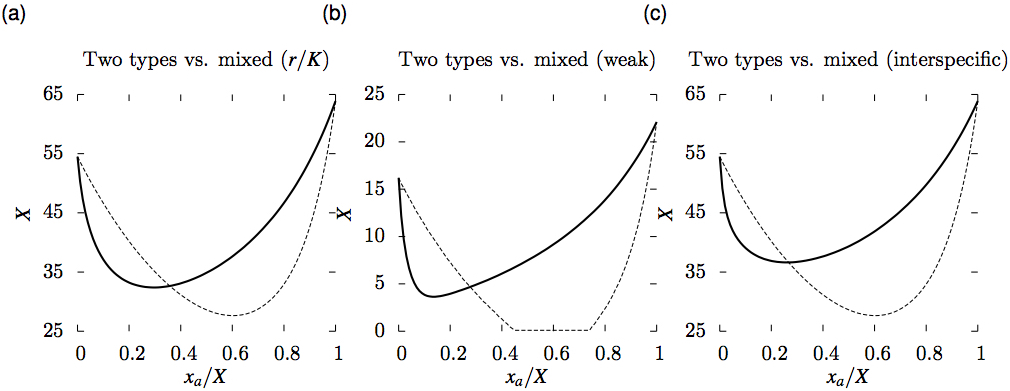}
\caption{Comparison of the separatrices of a polymorphic population of two types, altruists and cheaters, at initial condition $x_a/X$ (solid curves), and a monomorphic population with growth rate~\ref{average_Monod} (dashed curves). Below the curves, the immune system suppresses the pathogens, and above them, it does not. For low altruism, the two-type case outperforms the mixed type. And, for high altruism, the mixed type is optimal.} \label{compare}
\end{figure}

$K_\alpha$ is negatively correlated with cheater success (Figure~\ref{vary}). A low half velocity constant implies that the marginal benefit from the public good rapidly decreases as the proportion of altruists increases. As such, cheaters permit a more efficient utilization of the public good in the population. In the linear growth case, this effect cannot occur because the higher the proportion of altruists, the greater the total population's growth rate.

We compared the separatrices for the Monod models of a population of the two phenotypes and a population with a single phenotype with an intermediate production of the public good in Figure~\ref{compare}.  We plotted the population size required to overcome the immune system given an initial proportion of altruists $x_a/X$ for the two type case, and a single type with growth rate~\ref{average_Monod}.  When altruism is low, the two-type population is optimal for the pathogens. Conversely, when altruism is high, the mixed type population is optimal. If $x_a/X < \sqrt{\alpha K_\alpha/c} - K_\alpha$, then the mixed type will outcompete the two type case. Since, $x_a/X \to 0$ as $t \to \infty$ and~\ref{compare} is a decreasing function from $0$ to $\sqrt{\alpha K_\alpha/c} - K_\alpha$ in the two type case, while the growth rate of the mixed type will not decrease.

\section*{Discussion}

Previous studies support the hypothesis of frequency dependent selection among cheaters and altruists \cite{levin88,diggle07,ross07}. Altruists are less fit in the presence of cheaters, who outperform them. Further, average fitness is negatively correlated with the proportion of cheaters, which reduces virulence \cite{harrison06,rumbaugh09}. Our linear growth model qualitatively matches these empirical results. Given these observations, the question arises as to how altruism can be facilitated. However, less discussed, is why both cheating and altruism are prevalent, which is relevant since the prevalence of cheaters may be common \cite{velicer00,dugatkin05}.

Much research has explored mechanisms by which altruism can be facilitated, yet not how cheating can indirectly aide the population. Our approach was to explore how, in host-pathogen ecology, if cheaters may be necessary for pathogens to overcome the host's immune system. Our problem is essentially a threshold Volunteer's Dilemma \cite{diekmann85, archetti11}, where only if the population's public good production is sufficient, the group as a whole benefits. The Volunteer's Dilemma has been studied with respect to punishment \cite{raihani11}, shared rewards \cite{chen13}, voluntary reward funds \cite{sasaki14}, and asymmetric player strength \cite{he14}. Additionally, multilevel selection can favour a polymorphism of cooperators and defectors by maximizing the group donation level when the benefit function is sigmoid \cite{boza10}. We have shown that although cheaters out-compete altruists in a mixed population, such a population can be more virulent than a pure population of altruists or cheaters. This unexpected phenomenon occurs due to the Monod growth nature of the public good, as in the case of iron facilitated growth in \textit{P.\ aeruginosa} \cite{kim09}. The optimal total population growth rate may be at a mixed population. Although this harms altruists relative to cheaters, it may permit the pathogens to resist suppression by the immune system. Future work that studies the epidemiology of such a system would be interesting. A mixed population is required to overcome the immune response, during which the relative number of cheaters is increasing. However, if the carrying capacity for altruists were to be larger than for cheaters, than at large population sizes, the relative fitness advantage of cheaters may vanish. In this interplay, cheaters are $r$ selection and altruists $K$ selection phenotypes.

This research did not consider associativity amongst the bacteria, which may come about due to spatial effects. Future work employing associativity would be interesting, since the degree of associativity may have surprising effects. For example, high associativity between similar phenotypes would facilitate cooperation. However, this would reduce the optimization of the growth rate that occurs with cheaters in the Monod case.

\section*{Acknowledgements}

This material is based upon work supported by the Defense Advanced Research Projects Agency under Contract No. HR0011-16-C-0062.

\section*{Author contributions statement}

B.M. and D.C.V. conceived the theory and carried out the analytical work. B.M. ran the numerical simulations. Both authors wrote and reviewed the manuscript.

\section*{Additional information}

The authors declare no competing financial interests.

\nolinenumbers

\bibliography{bib_biofilm}
\bibliographystyle{abbrv}

\end{document}